**Recovery of Spectra of Phosphine in Venus' Clouds.**

Jane S. Greaves, Anita M. S. Richards, William Bains, Paul B. Rimmer, David L. Clements, Sara Seager, Janusz J. Petkowski, Clara Sousa-Silva, Sukrit Ranjan, Helen J. Fraser, on behalf of the authors of "Phosphine in the Cloud Decks of Venus" (Nature Astronomy, 14 September 2020; https://doi.org/10.1038/s41550-020-1174-4).

(In response to the *Matters Arising*: "Independent analyses find no evidence of phosphine in the atmosphere of Venus" by G. Villanueva et al.; their original *Matters Arising* entitled "No phosphine in the atmosphere of Venus" available at arXiv:2010.14305, with our original response entitled "Re-analysis of Phosphine in Venus' Clouds" at arXiv:2011.08176)

*Abstract:* We recover $PH_3$ in the atmosphere of Venus in data taken with ALMA, using three different calibration methods. The whole-planet signal is recovered with 5.4σ confidence using Venus bandpass self-calibration, and two simpler approaches are shown to yield example 4.5-4.8σ detections of the equatorial belt. Non-recovery by Villanueva et al. is attributable to (a) including areas of the planet with high spectral-artefacts and (b) retaining all antenna baselines which raises the noise by a factor ~2.5. We release a data-processing script that enables our whole-planet result to be reproduced.

The JCMT detection of $PH_3$ remains robust, with the alternative $SO_2$ attribution proposed by Villanueva et al. appearing inconsistent both in line-velocity and with millimetre-wavelength $SO_2$ monitoring. $SO_2$ contamination of the ALMA $PH_3$-line is minimal.

Net abundances for $PH_3$, in the gas column above ≈55 km, are up to ~20 ppb planet-wide with JCMT, and ~7 ppb with ALMA (but with signal-loss possible on scales approaching planetary size). Derived abundances will differ if $PH_3$ occupies restricted altitudes – molecules in the clouds will contribute significantly less absorption at line-centre than equivalent numbers of mesospheric molecules – but in the latter zone, $PH_3$ lifetime is expected to be short. Given we recover phosphine, we suggest possible solutions (requiring substantial further testing): a small collisional broadening coefficient could give narrow lines from lower altitude, or a high eddy diffusion coefficient could allow molecules to survive longer at higher altitudes. Alternatively, $PH_3$ could be actively produced by an unknown mechanism in the mesosphere, but this would need to be *in addition* to cloud-level $PH_3$ detected retrospectively by Pioneer-Venus.

*Response*

[We refer to our paper in Nature Astronomy as **G2020a**, and to the Matters Arising (second version) by Villanueva et al. as **V2021**; they refer to our first-version response as **G2020b**.]

We firstly find V2021's title "Independent analyses find no evidence of phosphine in the atmosphere of Venus" to be inappropriate. Non-recovery of the ALMA $PH_3$ line is attributable to higher noise levels in the V2021 analysis, as quantified below.

In their preliminary remarks, V2021 state "*The JCMT and ALMA data, as presented in G2020a/b are at spectral resolutions comparable to the frequency separation of the two lines. Moreover, the spectral features identified are several km/s in width, and therefore do not permit distinct spectroscopic separation of the candidate spectral lines of $PH_3$ and $SO_2$*". This is misleading, because the spectral resolution used for clarity in presentation is irrelevant to the line identification. Given that the intrinsic spectral resolutions of the datasets are very high (<0.1 km/s) the limiting factor is the noise level, which dictates the uncertainty in the calculated line-centroid velocity. G2020a/b quantified line centroids throughout, and here we find a centroid at -0.5 ± 0.6 km/s for our whole-planet spectrum. This is discrepant by 3σ with the potential contaminant $SO_2$ $30_{9,21}$-$31_{8,24}$, a feature which would appear offset by +1.3 km/s.

In their *Methods* section, V2021 now adopt the line broadening coefficient for $PH_3$ of G2020a.

In the following section, *ALMA Analysis*, V2021 state *"…large quasi-periodic fluctuations in the spectrum, which in G2020a/b is fitted with high-order polynomials. Particularly challenging is the fact that these fluctuations have a pattern/width comparable to their defined $PH_3$ line core region"*. This is misleading, because (a) two techniques are confused and (b) the fluctuations did not in fact have the same width. Regarding (a), G2020a adopted a high-order polynomial trendline to fit "ripply" spectra extracted from image cubes, but in G2020b and here, we do no such post-processing. Our bandpass self-calibration procedure does however apply, in the visibility domain, a 12$^{th}$-order amplitude solution (Supp. Fig. 1). This was recommended by an ALMA staff expert (not in our team) as a way to suppress ripples, and was tested on the simultaneously-observed HDO line (which V2021 agree is present). Regarding (b), widths of the original fluctuations were greater than that of the $PH_3$ line (full-width at half-minima, FWHM, of ~10 km/s versus 4-5 km/s – Supp. Fig. 4 and Table 1 in G2020a). We would **not** have argued that $PH_3$ was detected, if the absorption was simply one of a group of similar-width features.

V2021 then argue *"As we present in S2 and also reported in [12,13], artificially produced features which mimic true atmospheric lines can be produced when removing high-order polynomials from data with such characteristics"*. We comment that all three of these analyses have used some mathematically-incorrect orders for the polynomials they apply. G2020a described (on page 2, and Methods page 1) how a correct polynomial order is N+1, for N spectral "bumps" appearing within the passband and outside the interpolation region. However,

- V2021 fit a 12$^{th}$ order polynomial to a spectrum with only N=7 features (Figure FS2);
- ref. 12 (Figure 2) **over**-fit, with 12$^{th}$ order polynomials on passband sections with N≈3-5;
- ref. 13. (Figure 2) test **under**-fits, to JCMT N≈7 data using only 3$^{rd}$/4$^{th}$ order polynomials

and such approaches will intrinsically create additional or residual bumps in the spectra.

**ALMA-specific response**

We agree with V2021's remarks about the importance of the various calibration steps in the updated ALMA analyses, and thank them for notifying JAO of the "setJy" error. A clarification could be made to the caption of V2021's Figure 1, *"…re-analyzed data presented in G2020b using the new scripts released by JAO"*. In fact, ALMA notified groups known to be working on the data about the identified problems, and G2020b developed their own interim scripts.

Contrary to V2021, we detect $PH_3$, using multiple approaches to bandpass calibration, **including** ones based on the JAO re-released scripts. Our whole-planet $PH_3$ example is shown next (Figure 1). The Supplementary Information gives the details of all three calibration approaches we tested, along with additional figures showing $PH_3$ detections for smaller planetary areas. We also show in Figure 1 that $SO_2$ contamination of $PH_3$ is negligible; V2021 concur that $SO_2$ is not detectable in this ALMA dataset. Figure 1 additionally shows our whole-planet HDO detection, demonstrating similar HDO and $PH_3$ line profiles, and so robustness of the $PH_3$ detection.

Figure 1. *Bottom:* whole-planet ALMA spectrum of PH$_3$ (blue histogram) obtained from the Venus self-calibration bandpass solution and omitting antenna baselines < 33m. The zero-level (mean outside ±15 km/s) is ≈1 10$^{-5}$ Jy/beam; no polynomial fit was subtracted from the extracted spectrum. PH$_3$ line-area is 5.4σ over ±8 km/s. Line centroid is -0.5 ± 0.6 km/s. A constant-abundance model line-profile is overlaid (red dashes – see text). *Middle:* same PH$_3$ spectrum, overlaid (black line) with a ×0.2-scaled version of the HDO 2$_{2,0}$-3$_{1,3}$ spectrum, to demonstrate similar line profiles. HDO was observed simultaneously and reduced by the same method. *Top:* same PH$_3$ spectrum, overlaid with simultaneously-observed SO$_2$ 13$_{3,11}$–13$_{2,12}$ line, but scaled (from radiative transfer) ×0.024 and shifted by +1.3 km/s, to simulate the SO$_2$ 30$_{9,21}$–31$_{8,24}$ transition that overlaps with the PH$_3$ 1-0 line. The SO$_2$ spectrum was processed following the JAO QA3 scripts plus fitting of a 3$^{rd}$-order image-plane fit (see Supplementary Information).

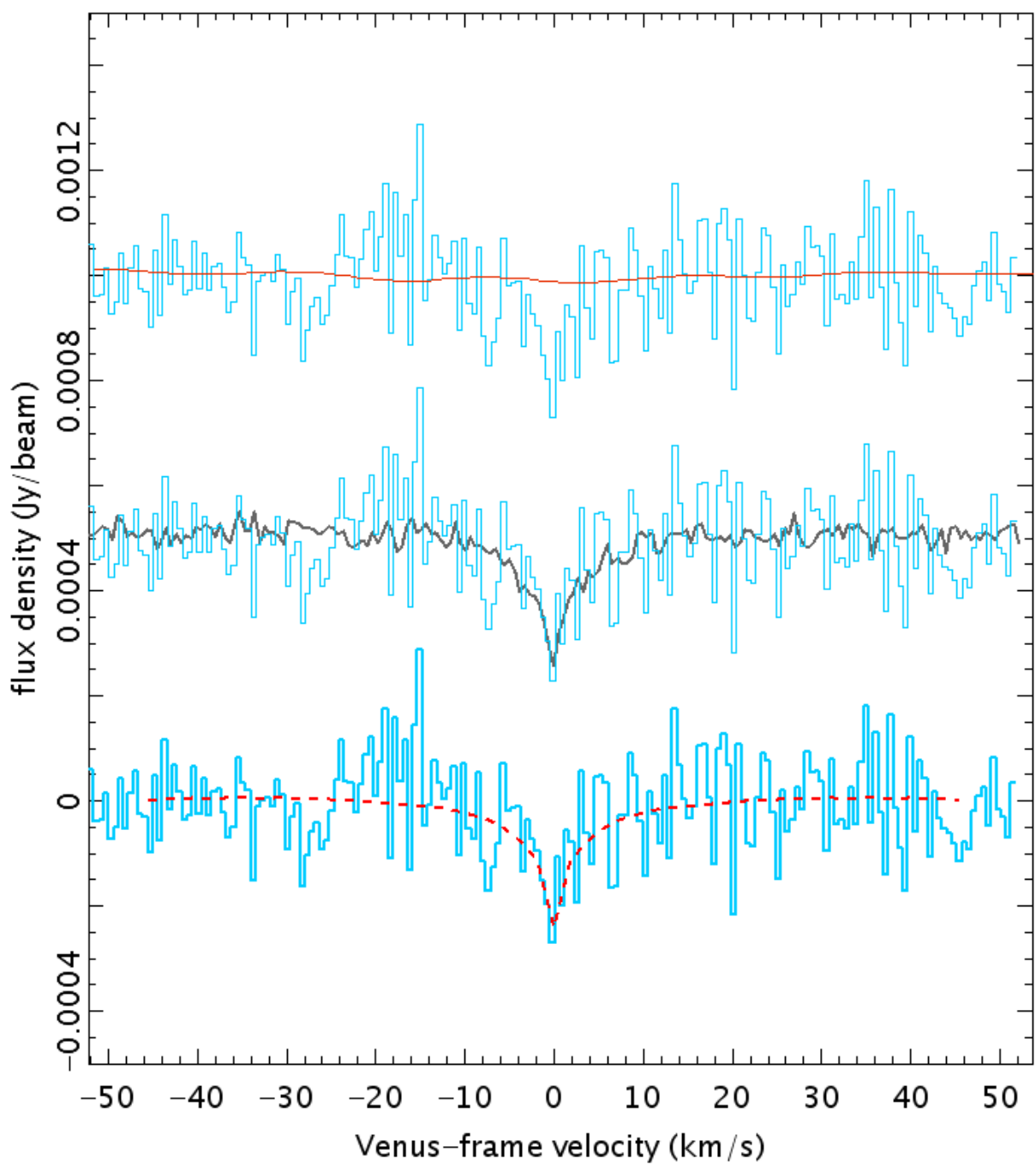

We show in Supplementary Figure 2 a further PH$_3$ detection, following methods that are closest to V2021's "JAO/Berkeley-CASA" analysis (their Figure 1D, spectrum 2). Both teams have here adopted the JAO-released bandpass calibrations. We found that, when imaged, the degree of residuals varies across the planet (Supp. Fig. 2, right panel) with upper/lower sections of the field still showing strong "ripple". However, excluding such areas yields good PH$_3$ detections, and Supplementary Figure 2 shows an example PH$_3$-recovery for Venus-latitudes within ±15º. This line-integrated detection is 4.8σ; further results are in preparation.

In particular, these factors have hindered V2021 in detecting phosphine:
- including the whole planetary area, as some regions have strong spectral ripples, and
- including all antenna baselines, as those of length <33m have worse ripples (G2020a).

Figure 2 illustrates how these factors can prohibit a detection. V2021's spectrum, including all antenna baselines, is significantly noisier than our spectrum that only includes baselines > 33m. While the ripples are similar in both analyses (peak-to-peak amplitude ≈0.275 mJy/beam), the V2021 channel-to-channel noise is higher by a factor ≈2.5. V2021's two alternate spectra appear similar (one slightly less and one slightly more noisy than the Figure 2 case).

Figure 2. *Top*: (y-offsets for clarity only) "JAO/Berkeley-CASA" whole planet spectrum, digitised from V2021 and converted to flux-density using G2020a's standard Venus model. The overlaid red-dashed curve is our radiative transfer model for PH$_3$, scaled for a fixed 1 ppb abundance, as a guide to detection limits. Bottom: our equivalent whole-planet spectrum following similar processing (Supplementary Information) and subtracting a 2$^{nd}$-order polynomial as in V2021.

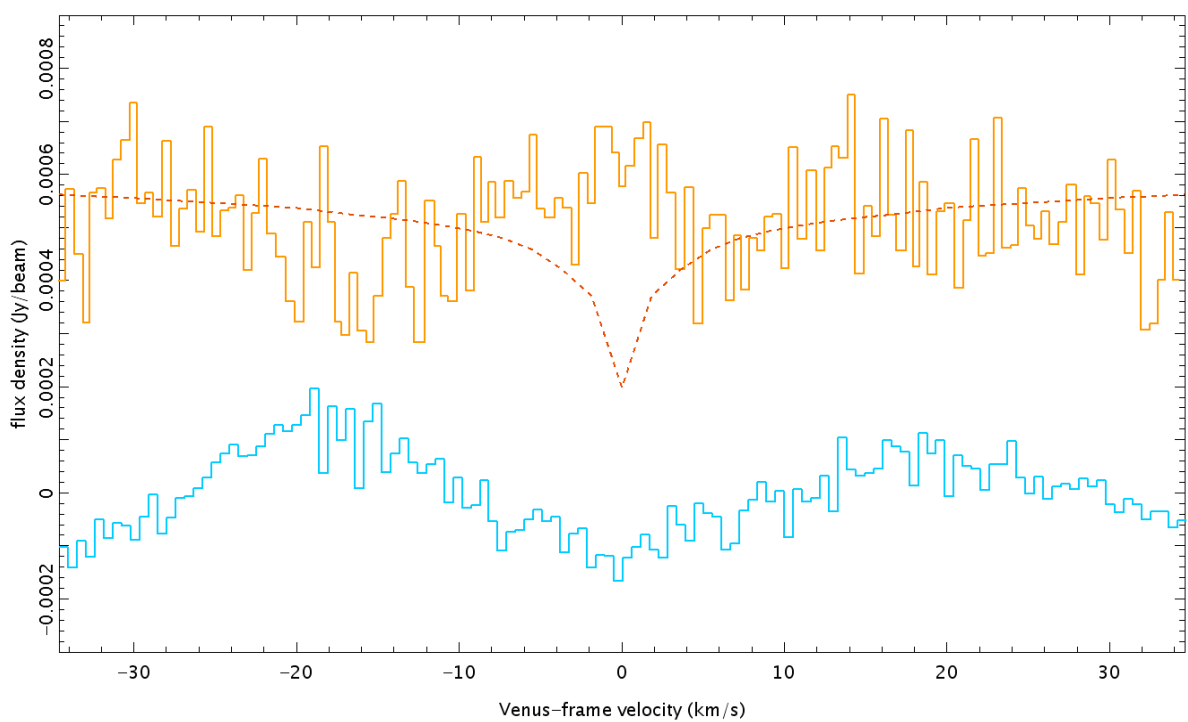

We agree that the **planet-wide** PH$_3$ signal can likely not be detected in the JAO-re-processed data, as some areas of the planet still exhibit large residuals (this was found even with some baselines omitted), and that ~1 ppb is a reasonable upper limit. However, higher abundances are inferred when some of the problematic data are excluded (Supplementary Information).

**JCMT-specific response**

In *JCMT analysis*, V2021 firstly comment on the 3-step process of removing spectral ripple – we make no further remark on this, as G2020a (Suppl. Info. pages 1-2) discussed the instrumental origins of the three separate effects.

V2021 then comment that ref. 13 "*explored the robustness of the two detection methods used in G2020a, namely low order polynomial fits and higher order multiple polynomial fits, and found that neither line detection method is able to recover a statistically significant detection at the position of the $PH_3/SO_2$ line*". This is a misunderstanding, as the high and low order fits are not alternatives, but are both necessary to combat different instrumental issues. Regarding statistical significance, ref. 13 used only a 1-parameter test, on line intensity, whereas G2020a relied on line-velocity as the second, more important, parameter in identification. In our *Addendum*, we show that with a 2-parameter test, the JCMT false-alarm probability is < ~1%.

V2021 then explore whether absorption **solely** by $SO_2$ could reproduce the JCMT line that G2020a interpreted as $PH_3$. V2021's approach has extreme freedom, as $SO_2$ exhibits high temporal variability plus complex behaviours over the altitudes traced by different techniques[1]. V2021 first apply an $SO_2$ altitude-profile from Ext. Data Fig. 9 of G2020a, although the figure was not intended by us for line-modelling (only to illustrate some constraints on photochemistry). V2021 then apply an altitude-profile that is mesospheric-heavy in $SO_2$ to model the JCMT spectrum. Issues here (their Figure 2) are that:

- any model $SO_2$-only line is intrinsically centred at +1.3 km/s, while G2020a's best-estimate for the line-centroid is -0.3 ± 0.9 km/s (1.8σ discrepancy);
- V2021's model line-width (FWHM ≈ 6 km/s) is difficult to reconcile with narrower model lines in other work (Figure 3; Figure 6 of Sandor et al.[2]; Figure 19 of Encrenaz et al.[3], Figure 2 of Lincowski et al.[4]);
- the "$PH_3$ Greaves+2020 (Fig. 9)" model mis-represents G2020a, as we would have adopted >20 ppb abundance for molecules placed only in the rarefied high atmosphere.

Figure 3. JCMT spectrum (G2020a, Figure 1b) overlaid with a radiative transfer model output (red dashes) scaled for 150 ppb of $SO_2$. A polynomial trendline was subtracted from the $SO_2$ model-line by the method applied to the observed data.

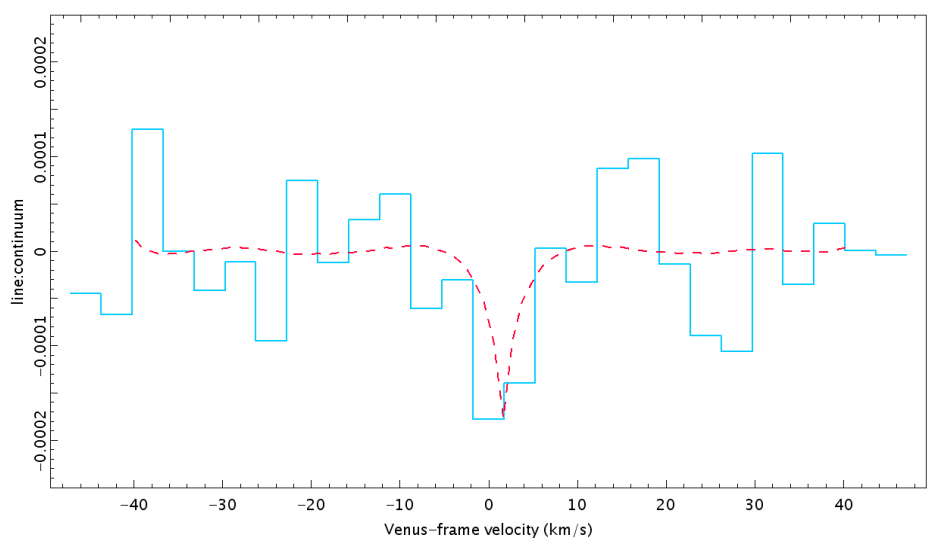

We demonstrate in Figure 3 that an exceptional $SO_2$ abundance would seemingly be needed to match the full line-depth observed with JCMT (i.e. to create a no-$PH_3$ line-solution). For comparison with extensive millimetre-wavelength monitoring[2], we used a constant abundance model, and estimate ~150 ppb of $SO_2$ would be required. Such an abundance would be a > +6σ outlier with respect to the monitored values[2], and in fact twice the maximum that was seen over several years. $SO_2$ would also need to have been **sustained** at this unusual level over the week of the JCMT observations.

As assumptions made about altitudes hosting sulphur dioxide can vary (V2021; ref. 2; Rimmer et al.[5]), Greaves et al. (in prep.) will investigate any resulting effects on the JCMT $SO_2$ line.

**Location of phosphine**

In *Probing altitude*, V2021 argue that $PH_3$ could only be detectable if it lies above 75 km altitude. We agree that the formation/survival of $PH_3$ at various altitudes is a problem for our current understanding of Venus' atmosphere. However, the simulations in V2021's Figure FS4 misrepresent our new ALMA reduction and our chemical models. V2021 place phosphine molecules in layers that are 10 km thick, simulate a spectrum, and then apply processing steps of subtracting a 6$^{th}$ order polynomial trendline while masking the central ±5 km/s. However,

- we did not apply such a polynomial trendline to the ALMA spectrum in G2020b, so the velocity-masking is not applicable (a span ~20 km/s could still be relevant: Supp. Fig. 1);
- V2021 do not show their model lines before subtracting polynomials, so we can not determine whether lower-altitude solutions could reproduce our spectra;
- modelling discrete 10 km-layers does not correctly represent the approach of G2020a, who inferred **one** particular layer of such thickness;
- it is incorrect to argue that lower altitudes do not contribute to the observed line – even though line-wings are lost in processing, this absorption still contributes to the line core.

As an model-free illustration of this last point, we consider altitude-ranges of 55-60 km (clouds) and 75-80 km (mesosphere). In a constant-abundance approximation, absorption in the cloud deck is favoured by a factor $~30\times(200/270)^{3/2}$ or ~20, from the relative numbers of molecules and temperature-dependence of the partition-function (e.g. Ext. Data Fig. 8 in G2020a). As the cloud-deck absorption will be spread over a line ~100× wider (for pressure ~300 mbar versus ~3 mbar in the mesospheric layer), the clouds' relative contribution to depth at line-core is ~20%.

We agree that some high-altitude molecules seem to be needed to produce a narrow line-core observed. One caveat is that a much smaller line-broadening coefficient could produce narrow lines at low altitudes – no altitude-retrieval can be robust until this coefficient is available.

We are presently revising our photochemical model. One possible approach is that $PH_3$ can survive above the cloud decks if the dynamic timescale is shorter than the chemical timescale. Figure 4 demonstrates an **ad-hoc** adjustment of the eddy diffusion coefficient $K_{zz}$ that can reproduce both the JCMT line and an IR upper-limit. We stress that this $K_{zz}$ would be unexpectedly high in relation to (sparse) in-situ measures, and might conflict with other results (e.g. the upper boundary $K_{zz}$ at the CO, $CO_2$ homopause). The model we use here is simplified to an atmosphere containing only $CO_2$ and $PH_3$; it adopts the profiles for *h*, *T*, *n*, *p*, *f*($PH_3$) of G2020a; and is then solved for intensity starting from *I* = 1 at 0 km, calculating Beer's law and accounting for pressure broadening and Doppler broadening (Voigt profile) at each height. This test model does not convolve with telescope beams as in G2020a, but demonstrates that an altitude-varying abundance chemical profile could generate a spectrum similar to the one observed. We agree with V2021 and ref. 4 that the original photochemical model of G2020a could not reproduce the observed line shape.

Figure 4. *Left*: ad-hoc altitude-profile with eddy diffusion coefficient $K_{zz}$ greatly increased in the mesosphere compared to the sparse in-situ constraints (Venera 9/10[6]; Pioneer[7]; purple shaded region is from Luginin et al.[8]). *Middle*: corresponding atmospheric profile of $PH_3$, compared to the G2020a profile. The red bar illustrates <5 ppb of $PH_3$ from 10 micron IR-observations[9]. The fixed 100 ppb of $PH_3$ at lower altitudes illustrates a possible abundance from Pioneer-Venus; the instrument-calibration description[10] suggests tens to hundreds ppb for the count rate extracted by Mogul et al.[11]. *Right*: the $PH_3$ 1-0 line that would be observed in this test model, along with the flat profile that the G2020a photochemistry would have produced.

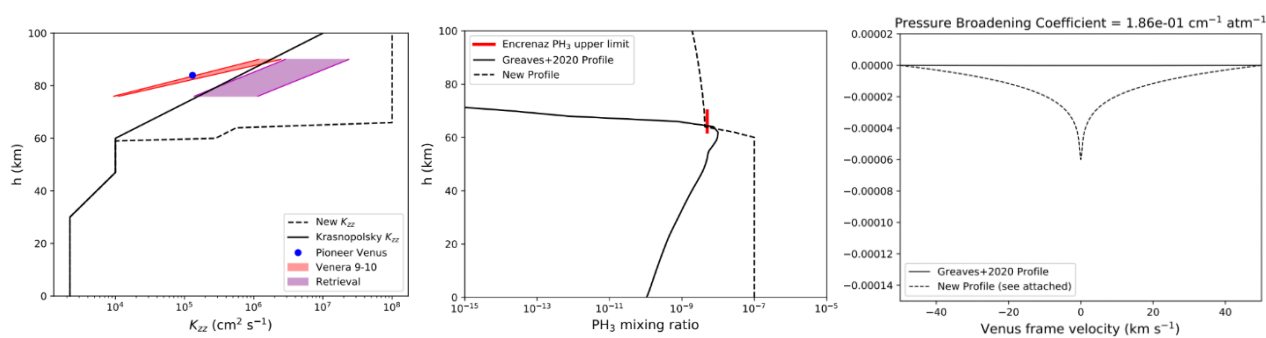

From this preliminary work, we consider that to produce the absorption observed: (1) $K_{zz}$ could be unexpectedly high above ~60 km, or (2) there could be an additional source of $PH_3$ in the mesosphere, or (3) $PH_3$ survives for a much longer time than chemical models thus far predict.

**Summary**

- $PH_3$ is recovered using three different methods of bandpass calibration, for various planetary areas;
- the independent ALMA analyses by V2021 do not reach sufficiently low noise for $PH_3$ to be detected (for the planet as a whole);
- V2021's $SO_2$-attribution for the JCMT line is anomalous with regard to line-centre velocity, line-widths from other models, and millimetre-monitored abundance;
- more work is needed to reconcile the $PH_3$ detections with radiative transfer and atmospheric chemistry models.

Finally, regarding the *Supplementary material* of V2021:

S1: We agree that $SO_2$ abundances at the time of JCMT observation in June 2017 do not constrain $SO_2$ abundances seen by ALMA in March 2019. We prefer abundances inferred from millimetre-monitoring[2] over IR spectra and in-situ measures (used in addition by V2021), owing to the difficulties in reconciling models of the different regimes[1].

S2: In summary, V2021's ALMA data-handling is quite similar to ours, with minor exceptions (such as V2021's use of a limb-darkened model of Venus). Our Supplementary Information presents the details of our data-processing. V2021 argue in their Figure FS1 that a 12$^{th}$ order polynomial post-processing of the spectrum based on JAO bandpass calibration can artificially produce a $PH_3$ line. However, this is a false analogy to our methods, since we now do not post-process extracted spectra with any polynomials. The 12$^{th}$-order fit to the **visibility bandpass** is discussed in Supplementary Information. Also, V2021's Figure FS2 is redundant, because the bandpasses shown – obtained by modifying G2020a's scripts and including all antenna baselines – are quasi-periodic with strong ripples. It is misleading to imply that our team would have generated a false detection by masking a spectrum of these characteristics, and so we **request that this figure be withdrawn or the accompanying text clarified**.

S3: V2021's validation on ALMA data produces very similar results to ours. We agree on an $SO_2$ abundance <10 ppb and an $H_2O$ abundance of ≈60 ppb (but in V2021's case for all-baseline data, and in our case for baselines > 33m only).

S4: The difficulties in interpreting the simulated spectra as a function of altitude in V2021's Figure FS4 were discussed above, in Location of phosphine.

## *Supplementary Information*

### *ALMA context*

Our project pushed ALMA's limits in detecting faint absorption against a strong continuum. The observations in March 2019 were highly time-constrained by issues of weather and antenna configuration, limiting calibrator choices, with Jupiter's moon Callisto chosen as the brightest available bandpass-calibrator. Using an ephemeris-object required a full model to be inserted during calibration, but this is not needed for the QSOs normally used and was not standard practice for standard "QA2" calibration in 2019. This caused the model to be used incorrectly, both in setting the flux scale (too high) and more significantly in introducing severe bandpass ripples. These aspects were not known to the science team until after publication of G2020a.

A critical step in G2020a was a de-trending procedure to characterise ripple in the bandpass. G2020a used a mathematically-correct 12$^{th}$-order polynomial, **solely** in order to demonstrate a wide section of the bandpass. G020a's Table 1 also showed equivalent results with a 1$^{st}$-order fit, and Ext. Data Fig. 4 demonstrated that the 12$^{th}$-order polynomial did not create "fake lines".

However, this type of de-trending relies on the ripple superimposed on any real absorption line having similar characteristics to ripples elsewhere in the bandpass. This assumption now appears invalid, for two possible reasons:

- in standard calibration, the bandpass calibrator data are channel-averaged (over 4 MHz in G2020a), to improve signal to noise – but it has now been shown by the observatory that this could introduce a "zigzag" at low levels in line:continuum (below ~1:1000)

and in addition it is possible that:

- the centre of the passband can show peculiarities if the frequency of interest is tuned to the passband centre in simultaneous narrowband and wideband settings, as in G2020a.

The intent is to re-observe Venus using optimised ALMA settings. This could include longer calibrations, using a small mosaic if Venus fills the primary beam, and applying a higher-precision primary beam model in order to extract faint absorption lines from small areas. For now, the ALMA data gathered in March 2019 have contributed to better processing of lines at very high dynamic range. Data re-processing was developed and led by the ESO ALMA Regional Centre with products now available in the JAO public archive. Other datasets affected by similar bandpass calibration issues are also being re-processed and re-released by the observatory.

*Re-processing*

ESO have released a full description (see the updated archive README and script notes) of the new processing steps that have been identified. Along with some minor updates (such as to antenna position tables), the major factors are:

- corrections to the flux scale (previously, the effect of Venus on system temperature was not handled correctly by standard procedures);
- including a time-dependent model for Callisto (implemented via usescratch=True in the CASA task setJy);
- optimising the bandpass correction for Callisto, including using smooth polynomial fitting in the CASA bandpass task;
- using a more recent version of CASA (the one available in 2019 applied the primary beam correction wrongly to ephemeris objects).

The JAO bandpass solution does not completely remove both bandpass "bumps" and spatial "ripples". Supplementary Figure 1 compares bumps remaining in the Venus spectrum after applying only bandpass calibration derived from Callisto, and after more advanced processing. This approach – suggested to us by independent observatory staff – runs a bandpass self-calibration on Venus itself. The demonstration uses the HDO data (a good test, as the expected line-shape is known) and shows the result of a Venus bandpass self-calibration with a $12^{th}$-order polynomial for amplitude and a $5^{th}$-order polynomial for phase. We note that the success of this method **supports the subtraction of $12^{th}$-order polynomials** by G2020a; i.e. G2020a correctly identified the approximate order of bandpass characteristic. However, solutions are now applied in the visibility domain, which helps to remove subjective biases about how spectra should look after imaging and extraction.

Supplementary Figure 1 HDO $2_{2,0}$-$3_{1,3}$ data, (*left*) before any bandpass self-calibration and (*right*) after Venus bandpass self-calibration as described in the text, under Method (i).

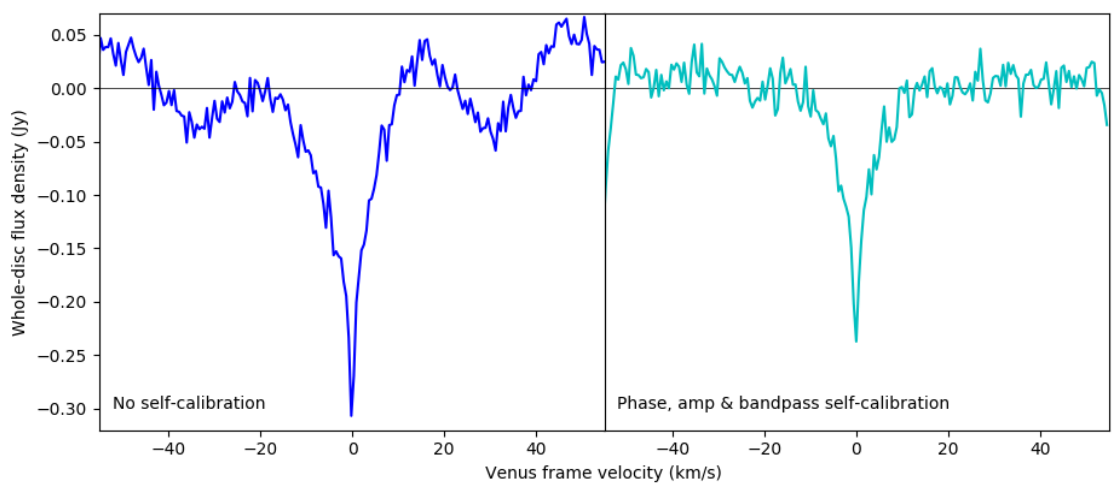

We eventually tested three methods of bandpass calibration. All of these supersede the data products in G2020a, and so abundance and latitude analyses will be revised (in a *Correction*).

(i) The approach with bandpass self-calibration using Venus itself is demonstrated in Supp. Fig. 1, with the resulting whole-planet $PH_3$ and HDO lines shown in Figure 1. We followed QA3 calibration up to application to Venus, excluded baselines < 33m, and used a linear fit at the uvcontsub stage. The Venus data were then phase and amplitude self-calibrated (via the Butler-JPL-Horizons model) and then bandpass self-calibrated using a $12^{th}$-order amplitude, $5^{th}$-order phase polynomial, before continuum subtraction. No post-processing polynomials were fitted to extracted spectra. The left plot in Supp. Fig. 1 was extracted from a data-cube made similarly but before any self-calibration; there are <12 nodes visible as the band edges were not imaged. This image was scaled to the Butler-JPL-Horizons 2012 flux density by multiplication, comparing the all-baseline continuum disc flux density to that of the model.

The HDO example (Supp. Fig. 1, right) shows that most of the spectral ripple has been removed, and thus the same procedure was applied to the $PH_3$ data. A small amount of the line fluxes may have been removed. Shallower bandpass ripples are seen than in G2020a, but some spatial gradients and spectral artefacts remain. Ripples are still worse on shorter baselines. The noise is slightly increased compared to our Method (ii), probably due to amplitude aspects of self-calibration being affected by the interferometric nulls.

(ii) The data as re-archived by JAO ("QA3") adopted a Callisto bandpass calibration solution that applied a polynomial of $3^{rd}$-order in amplitude and $5^{th}$-order in phase. In our Method (ii), there is no subsequent bandpass self-calibration. We ran scripts as for QA3 but omitted amplitude self-calibration on Venus (V2021's choice also for their methods 2, 3). We then subtracted the continuum using a linear fit to the visibilities, excluding only edge-channels, and imaged each spectral line separately in continuum-uv-subtracted data. Antenna baselines <33m were again omitted. We then tested low-order fits in the image plane, to remove the need for post-imaging trendline-fitting. We found the lowest approximately-optimal order to be 6 for the narrowband $PH_3$ and HDO data; orders are now lower in part because edge channels with greatest deviations were excluded from imaging. Orders >3 did not significantly improve the problematic wideband data (covering $SO_2$ $13_{3,11}$–$13_{2,12}$). These image-plane fits were subtracted from the datacubes. No post-processing polynomial fits were applied to extracted spectra.

In this method, we found that some regions of the imaged plane still exhibit strong spectral ripple. The right panel of Supplementary Figure 2 illustrates spectra from 8 horizontal strips across Venus, showing that including problematic upper/lower sections of the field will tend to overwhelm any $PH_3$ line signal. A $PH_3$ detection is demonstrated for the equatorial belt of Venus, i.e. excluding the most "ripply" areas (Supplementary Figure 2). The line-area integrated over ±10 km/s yields a 4.8σ detection, after a small zero-level correction.

<u>Supplementary Figure 2.</u> *Left*: net $PH_3$ 1-0 spectrum from Method (ii) for ±15° latitude range (blue histogram). The overlaid orange dashed curve is a Voigt profile (fitted within ±15 km/s) as a guide and to illustrate the estimated zero-level. The inset plot compares the equivalent-area in the G2020a data (grey histogram: no polynomial fitted, pre-corrections flux-density scale). *Right*: Method (ii) spectra extracted from adjoining strips that span Venus horizontally (in RA) and have vertical heights of 12 pixels (1.92 arcsec). Spectral binning is used for clarity. Plots are organised (bottom to top) from nearest the image centre to the Declination extremes, with +Dec. regions shown as histograms and -Dec. regions as line plots. Venus' equator was tilted by 13° with respect to the RA axis – spectral extraction in the RA,Dec. frame shows instrumental effects more clearly than similar strips oriented parallel to the equator.

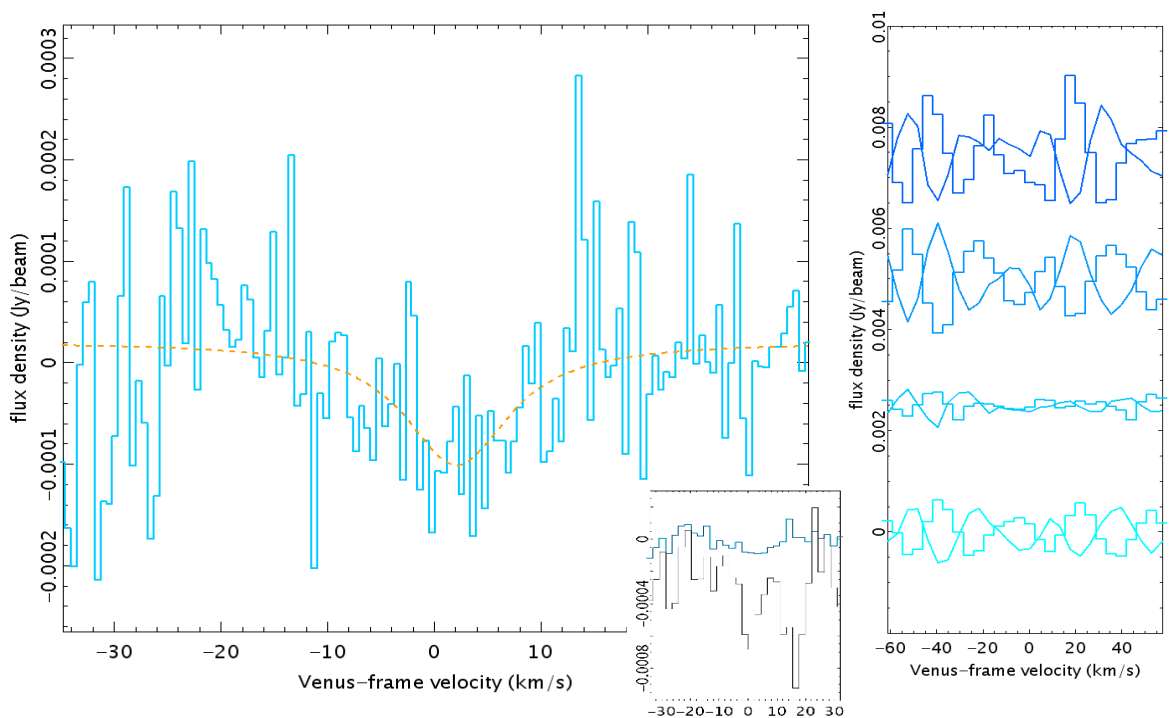

(iii) A final "minimal-intervention" calibration method was tested (G2020b), where the Callisto conventionally-averaged bandpass was simply smoothed and then applied to Venus. We show an example PH$_3$ detection in Supplementary Figure 3, for the ±15° equatorial zone as in Supplementary Figure 2 (but here for only 92% of the planetary diameter, to mitigate against ripple effects at the limb). The line-cores are recovered for both HDO and PH$_3$, but with only a few spectral channels detected, so we did not use this method further. For PH$_3$, there is a 4.5σ detection in the line-centre channel, while the same channel is 4.9σ in the HDO spectrum.

Supplementary Figure 3. Example results from the smoothed Callisto-bandpass approach. The region sampled is the ±15°-latitude equatorial band, cropped from the limbs to 92% planetary diameter. The lower spectrum is PH$_3$ 1-0 and the upper offset spectrum is HDO 2$_{2,0}$-3$_{1,3}$.

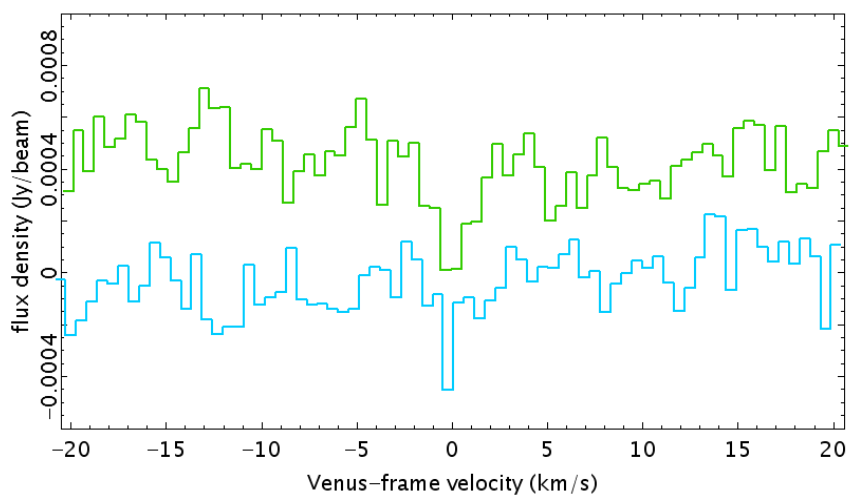

*Abundances*

Inferred abundances depend to some extent on adopted flux-scaling. A bright target such as Venus contributes to the system temperatures used to convert raw signals to flux, while amplitude self-calibration is difficult for bright objects with sharp edges (as discussed by V2021). For this reason, G2020a used a model of Venus-flux to establish line:continuum (l:c) ratios. Here we use our in-parallel continuum-flux measurements to estimate abundances over particular angular scales.

The verification spectra of HDO $2_{2,0}$-$3_{1,3}$, agree extremely well between our analysis and that of V2021. V2021 fit their HDO whole-planet detection with a model of 60 ppb of $H_2O$ (folding in D/H = 200 above terrestrial), which produces a line-minimum l:c of -2 $10^{-4}$ (their Figure FS3). We have re-processed the HDO data starting from the QA3 script, as in our Method (ii), but for test purposes retaining all antenna baselines. We then find l:c at line minimum of -2.0 $10^{-4}$, precisely matching the V2021 model. We note that if antenna baselines < 33m are omitted, l:c$_{min}$ is doubled, to -4.0 $10^{-4}$. This suggests that the continuum is more smoothly distributed than water vapour, i.e. relatively more continuum signal is lost when sampling only smaller scales (for baselines > 33m: only up to 4.3 arcsec, or ~30% planetary-diameter).

For our whole-planet $PH_3$ detection using Venus bandpass self-calibration (Figure 1), we find l:c$_{min}$ of -1.8 $10^{-4}$. This applies to molecule-distributions of size up to ~30% planetary-diameter, and includes only the 22% of the continuum signal captured on the long antenna baselines (this percentage was checked from "QA3"-based processing). For our constant-abundance radiative-transfer model line (G2020a), we then flattened the wings outside ±10 km/s, to simulate suppression of wings in the data during bandpass calibration (Supp. Fig. 1). The best line match for the whole-planet spectrum (Figure 1) is then ~7 ppb of $PH_3$. We emphasise that (a) this abundance could change depending on the actual spatial distribution (as for HDO), and (b) no reconciliation with the photochemistry is as yet attempted. It is as yet unclear which of effects (a) and (b) could most alter the derived abundance (see V2021 and Akins et al.[12]).

By the same method of in-parallel continuum measurement, we find line-minimum l:c of -1.8 $10^{-4}$ for the equatorial zone in the JAO-basis processing (Supp. Fig. 2). Hence the nominal equatorial abundance will be very similar to the nominal whole-planet value.

Using the same radiative transfer model, G2020a estimated 20 ppb for the JCMT detection of $PH_3$; this would be reduced if there is some $SO_2$ contamination. Our new estimate of ~7 ppb with ALMA is in reasonable agreement (but only samples scales up to ~30% planet-diameter). These values could connect independent results of < 5 ppb above 60 km[9] and tens-to-hundreds ppb at 51 km[10,11]. However, our derived phosphine abundance and that of ref. 9 may alter when improved physical coefficients become available.

*Acknowledgements*: The G2020 team thank the many ALMA staff who contributed tirelessly and speedily to this re-processing project, developing new tests and techniques in a very short time period. The work was led from ESO with input from JAO and NAASC.

*Scripts*: to be supplied on publication.

[1] Shao, W. D., Zhang, X., Bierson, C. J., & Encrenaz, T. Revisiting the Sulfur-Water Chemical System in the Middle Atmosphere of Venus. Journal of Geophysical Research (Planets) 125, 06195 (2020).

[2] Sandor, B. J. & Clancy, R. T. First measurements of ClO in the Venus atmosphere–altitude dependence and temporal variation. Icarus 313, 15–24 (2018).

[3] Encrenaz, T., Moreno, R., Moullet, A., Lellouch, E. & Fouchet, T. Submillimeter mapping of mesospheric minor species on Venus with ALMA. Planet. Space Sci. 113, 275–291 (2015)

[4] Lincowski, A. P. et al. Claimed Detection of $PH_3$ in the Clouds of Venus Is Consistent with Mesospheric $SO_2$. Astrophysical Journal Letters 908, L44-52 (2021).

[5] Rimmer, P. B. et al., Three Different Ways to Explain the Sulfur Depletion in the Clouds of Venus. Planetary Science Journal, submitted; arXiv:2101.08582 (2021).

[6] Krasnopolsky, V., in Hunten, D.M. ed., Venus (Space Science Series). University of Arizona Press (1983).

[7] Lane, W.A. & Opstbaum, R., High altitude Venus haze from Pioneer Venus limb scans. Icarus 54, 48-58 (1983).

[8] Luginin, M., Fedorova, A., Belyaev, D., Montmessin, F., Korablev, O. & Bertaux, J.L. Scale heights and detached haze layers in the mesosphere of Venus from SPICAV IR data. Icarus 311, 87-104 (2018).

[9] Encrenaz, T. et al. A stringent upper limit of the $PH_3$ abundance at the cloud top of Venus. Astronomy & Astrophysics 643, L5 (2020).

[10] Hoffman, J. H.; Hodges, R. R.; Donahue, T. M.; McElroy, M. M. Composition of the Venus lower atmosphere from the Pioneer Venus mass spectrometer. Journal of Geophysical Research 85, 7882-7890 (1980).

[11] Mogul, R., Limaye, S. S., Way, M. J. & Cordova, J. A. Venus' Mass Spectra Show Signs of Disequilibria in the Middle Clouds. Geophysical Research Letters 48, e91327 (2021).

[12] Akins, A. B., Lincowski, A. P., Meadows, V. S. & Steffes, P. G. Complications in the ALMA Detection of Phosphine at Venus. Astrophysical Journal Letters 907, L27-33 (2021).